\documentclass{book}    
\usepackage{piers}  
\usepackage[]{graphicx}
\graphicspath{{Figures/}}
\usepackage{subfigure}
\usepackage{amsmath}
\usepackage{amssymb}
\usepackage{epsfig}
\usepackage{cite}
\usepackage{color}
\usepackage{balance}
\usepackage{pgfplots}
\usepackage{wrapfig}
\usepackage{lipsum}
\usepackage{hyperref}

\begin{document}

\title{A Chaotic Approach on Solar Irradiance Forecasting}
\maketitle

\author      {F. M. Lastname}
\affiliation {University}
\address     {}
\city        {Boston}
\postalcode  {}
\country     {USA}
\phone       {345566}    
\fax         {233445}    
\email       {email@email.com}  
\misc        { }  
\nomakeauthor

\author      {F. M. Lastname}
\affiliation {University}
\address     {}
\city        {Boston}
\postalcode  {}
\country     {USA}
\phone       {345566}    
\fax         {233445}    
\email       {email@email.com}  
\misc        { }  
\nomakeauthor

\begin{authors}

{\bf T. A. Fathima}$^{1}$, {\bf Vasudevan Nedumpozhimana}$^{2}$, {\bf Yee Hui Lee}$^{3}$, \\{\bf Stefan Winkler}$^{4}$, {\bf and Soumyabrata Dev}$^{2,5,6*}$\\
\medskip

$^{1}$Indian Institute of Technology Bombay, Powai, Mumbai, India\\

$^{2}$ADAPT SFI Research Centre, Dublin, Ireland\\

$^{3}$Nanyang Technological University Singapore, Singapore 639798\\

$^{4}$ Department of Computer Science, National University of Singapore, Singapore 117417\\

$^{5}$School of Computer Science, University College Dublin, Ireland.

$^{6}$UCD Earth Institute, University College Dublin, Ireland.

$^{*}$ Presenting author and corresponding author

\end{authors}

\begin{paper}

\begin{piersabstract}
We analyse the time series of solar irradiance measurements using chaos theory. The False Nearest Neighbour method (FNN), one of the most common methods of chaotic analysis is used for the analysis. One year data from the weather station located at Nanyang Technological University (NTU) Singapore with a temporal resolution of $1$ minute is employed for the study. The data is sampled at $60$ minutes interval and $30$ minutes interval for the analysis using the FNN method. Our experiments revealed that the optimum dimension required for solar irradiance is $4$ for both samplings. This indicates that a minimum of $4$ dimensions is required for embedding the data for the best representation of input. This study on obtaining the embedding dimension of solar irradiance measurement will greatly assist in fixing the number of previous data required for solar irradiance forecasting. 
\end{piersabstract}

\psection{Introduction}
The time series modelling can be easier once the behaviour of time series is well studied. It helps to find out any hidden determinism underlying in the time series data. The analysis of time series data will be helpful to estimate the minimum number of dimensions (variables) required to predict its future state. The time series of most of the natural processes looks erratic~\cite{manandhar2019predicting}, however according to chaos theory; these complex patterns can have simple causes. This concept has motivated to apply it for analysing the time series data of solar irradiance, a regular event that occurs in the atmosphere. Out of the various methods available to study the chaotic behaviour of a time series, False Nearest Neighbour method (FNN) is used in the present study.

Chaos theory was developed to understand the dynamical systems with randomness over time in a better way. Lorenz~\cite{lorenz-1963} was the first one who reported chaotic characteristics of atmosphere and climate while developing numerical models. In general, chaos theory is the study of forever changing complex systems (dynamical systems) based on mathematical concepts of recursion~\cite{williams1997}. The change and time are the two subjects which together constitute the foundation of chaos. Even though a system is dynamic and showing random behaviour, chaos theory opens a realm to understand the static property within a dynamical complex system. The behavioural analysis of a time series is essential to understand the stable properties in its complex evolution. The embedding dimension is a way to mathematically characterize the `complexity' of the system’s evolution which can be used for better representation of input data.

Better representation of data is very essential prior to the development of any learning models. False nearest neighbour method is the most widely used tool to estimate the minimum embedding dimension needed to unfold the behaviour of a time series. The present study is focused to estimate the optimal embedding dimension of solar irradiance of Singapore weather data with the help of false nearest neighbour method. We performed experiments on one year of time series data with a frequency of $60$ minutes samples and $30$ minutes samples. 

\psection{Chaos Theory on Solar Irradiance}

Solar irradiance is the power per unit area of electromagnetic radiation received from sun. It is measured in W/m$^2$.  The analysis of solar irradiance and its prediction is important, as it has many applications, especially in the energy generation for solar based power plants. The solar irradiance also plays a role in climate modelling and weather forecasting. Medvigy and Beaulieu~\cite{medvigy2012} observed a correlation between solar irradiance and precipitation. Researchers started using this variable for predicting the onset of precipitation~\cite{manandhar2019}. However, the variation of solar irradiance is quite erratic (chaotic nature) due to atmospheric conditions like cloudiness which makes its prediction difficult. Hence, proper analysis of solar irradiance time series data is crucial. The false nearest neighbours (FNN) algorithm is used as a method for determining the proper embedding dimension for recreating or unfolding the dynamics of non-linear systems \cite{kennel1992,abarbanel1993}.

For the sake of generality and a complete understanding of chaos theory, we explain the important terms that are used throughout this paper. 
Let us suppose that $x_t$ and $x_{t+1}$ are the solar irradiance measurements at time instants $t$ and $t+1$. 

\psubsection{Phase space}

A standard phase space is the graph between different kinds of variables. However, in chaos theory, mostly pseudo phase space diagram is used for the behavioural study. Pseudo phase space diagram is an imaginary graphical space in which the axes represent values of one physical feature,taken at different times. The pseudo space co-ordinates represent the variables needed to specify the state of a dynamical system. Some of the time series can be very long and therefore difficult to show on a single graph, a phase space plot condenses all the data into a manageable space on a graph. Although no one can draw a graph for more than three variables while still keeping the axes at right angles to one another, the idea of phase space holds for any number of variables~\cite{williams1997}. The deterministic time series confines a small area in the phase space plot but for stochastic time series the phase points may fill the entire space~\cite{wei2010}.

\psubsection{Lag}

In this work, the method of time-delayed embedding is used for reconstructing a higher dimensional pseudo phase-space from a one dimensional time series \cite{Packard1980,Takens1981}. Pseudo phase space is a setting for comparing a time series to later measurements within the same data (a sub-series). For instance, a plot of $x_{t+1}$ vs $x_t$ shows how each observation ($x_t$) compares to the next one ($x_{t+1}$). In that comparison, the group of $x_t$ values represent the basic series and the group of $x_{t+1}$ values represent the sub-series. By extending that idea, each observation ($x_t$) can be compared with two later measurements or sub-series ($x_{t+1}$ and $x_{t+2}$) and so on. Lag is selected as a constant interval in time between the basic series and any other sub-series. It specifies the rule or basis for defining the sub-series. For instance, when compared with the basic series ($x_t$), the sub-series $x_{t+1}$ is with a lag of one, sub-series $x_{t+2}$ is with a lag of two and so on. Lagged phase space is a special type of pseudo phase space in which the coordinates represent lagged values of one physical feature. Such a graph is long-established and common in time-series analysis. It is a basic, important, and standard tool in chaos theory as well~\cite{williams1997}.

\psubsection{Embedding dimension\label{embedding-dim}}

The embedding dimension is the total number of separate time series (including the original series, plus the shorter series obtained by lagging that series) included in chaotic analysis. The analysis need not involve every possible sub-series. For example, one might set the lag at one, and compare $x_t$ with $x_{t+1}$, in that case the embedding dimension is two. The next step might be to add another sub-series $x_{t+2}$ and consider groups $x_t$, $x_{t+1}$, and $x_{t+2}$ together, in that case embedding dimension is three and so on. From a graphical point of view, the embedding dimension is the number of axes on a pseudo phase space graph. Analytically, it is the number of series used in the chaotic analysis. For prediction purposes, we can construct pseudo phase space of dimension $m$ which can be decided using FNN method. From a prediction point of view, the embedding dimension is the number of previous inputs that is used to predict the future value in the series plus one. Thus embedding dimension gives an idea to use the number of previous inputs required for predicting the future state of the variable.

\psection{False Nearest Neighbour Method}

The false nearest neighbour method (FNN) is one of the effective  method for estimating the embedding dimension from the phase space reconstruction\cite{Wu2010,Ghorbani2012,Sebastian2018,albostan2015}. Adenan and Noorani\cite{Adenan2013} developed a data driven prediction model for stream flow time series using the embedding dimension obtained from FNN algorithm and reported a better prediction result.The FNN algorithm is based on the geometry of the phase space reconstructed using a single time series. If there is enough information in the reconstructed phase space to predict the future output, then two vectors which are close in the reconstructed phase space will also have future outputs which are close~\cite{kennel1992}. If the embedding is not done in enough dimensions then there will be some neighbours in the space with vastly different outputs, even though their trajectories are close in the phase space. Such neighbours are called false neighbours since they are close because of projection onto a space with a dimension too small to represent the dynamics of the system. This is illustrated graphically in Fig.~\ref{fig:fnn}. From this figure, it is clear that phase space vectors $P_1$ and $P_2$ are not nearest neighbours to each other. The euclidean distance between these two vectors are calculated in one dimension ($d_1$) as well as in two dimensions ($d_2$). However, due to projection in one dimension (insufficient to represent actual geometry), they appear to be close neighbours.

\begin{figure}[htb]
    \centering
    \includegraphics{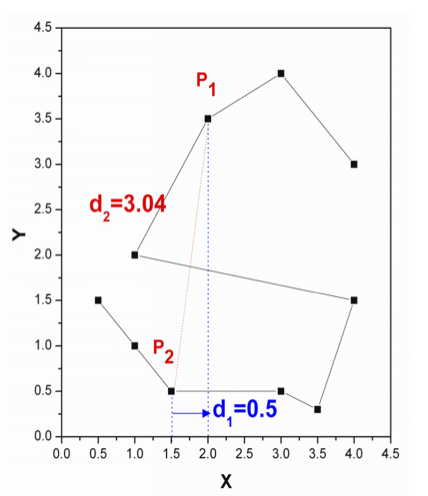}
    \caption{Graphical illustration of false neighbours.}
    \label{fig:fnn}
\end{figure}

In order to find whether the neighbours are true or false, a ratio test is conducted to determine whether the distance between the future values are significantly larger than the distance between the neighbours which are close in reconstructed phase space. In FNN method for each vector in the phase space $Y_i$, its nearest neighbours $Y_j$  are searched in the $m$ dimensional phase space. The euclidean distance ($d_m$) between $Y_i$ and $Y_j$ is calculated in the dimension $m$. Then, their euclidean distance ($d_{m+1}$) in $(m+1)^{th}$ dimension is estimated and compared with $d_m$. If the ratio between those two distances ($d_{m+1}/d_m$) is greater than a threshold value, then the vectors are considered as false neighbours.

This is carried out for each and every vector in phase space and percentage of vectors in the data which have false neighbours is found. This
algorithm is repeated for higher dimensions (m) until the percentage of false neighbours drops to acceptably small number. Then the dimension of the reconstructed vector is said to be large enough to make accurate prediction of future outputs. Thus, for chaotic time series the fraction of false neighbours decreases as the embedding dimension increases. This happens because the closer vectors in lower dimension remain closer to each other in higher dimensions. Until the proper embedding is reached, the fraction of false neighbours will be high and it decreases gradually. The advantages of using FNN method for finding the dimension of the system are: (i) it does not contain any subjective parameters except for the time-delay for the embedding; (ii) it does not strongly depend on how many data points are available; and (iii) is an easier method to implement in practical applications~\cite{Cao1997}. The FNN method can suggest the acceptable minimum embedding dimension required by looking at the behaviour of neighbours when the embedding dimension changes from $m$ to $m+1$~\cite{kennel1992}. In reconstructed phase space, the false neighbours are the vectors which lie in the neighbourhood of other vectors, but they do not belong to the neighbourhood of same vectors when projected to higher dimensions.

\psubsection{Step by Step Procedure}
\begin{enumerate}
    \item The dynamics of time series ($x_1 ,x_2 ,\dots x_n$) are fully captured or embedded in the m-dimensional phase space as shown below
    \begin{equation*}
        (x_1,x_2,\dots)\longrightarrow \{(x_1,x_2,\dots,x_m),(x_2,x_3,\dots,x_{m+1}),\dots\}
    \label{eqn:somelabel}
    \end{equation*}
    \item Identify the nearest vector to a given vector in the phase space for a dimension $m$ based on the euclidean distance.
    \item Determine the ratio $d_{m+1}/d_m$ for the same pair of vectors in both the dimensions $m$ and $m+1$.
    \item Check whether the ratio is greater than the threshold value. If it is greater, the neighbour is considered as false.
    \item Repeat the procedure with all the vectors in phase space and find the percentage of false neighbours for each embedding dimension by comparing with one higher dimension
    \item Increase the embedding dimension and repeat the procedure until the percentage of false neighbour drops to a lower value which will be the dimension required for the prediction model.
\end{enumerate}

\psection{Experiments and Results}
In our experiments, we used 1 year solar irradiance data from the weather station located at Nanyang Technological University (NTU). This solar irradiance data have temporal resolution of 1 minute. From this data, two data samples with temporal resolution of 60 minutes (solar-60min) and 30 minutes (solar-30min)are prepared for the chaotic analysis. The solar irradiance data with $60$ minute resolution (solar-60min) had $8760$ samples with mean $160.68$ and standard deviation $255.57$. The solar irradiance data with 30 minutes resolution (solar-30min) had $17520$ samples with mean value $161.05$ and standard deviation $255.42$. Both solar-30min and solar-60min shows similar statistical behaviour. 

False Nearest Neighbour method is applied on both solar-60min and solar-30min data to measure percentage of false neighbours. Percentage of false neighbours for varying embedding dimensions from 1 to 7 are plotted in Fig.~\ref{fig:fnn-result}. From the plot, it can be seen that for both solar-60min and solar-30min, the percentage of false neighbours is saturated to its minimum value at an embedding dimension of 4.

\begin{figure}[htb]
\centering
\begin{tikzpicture}
\begin{axis}[xlabel={Embedding dimension}, ylabel={\% false neighbours}, xmin=1, xmax=7, ymin=0, ymax=100,xtick={0,1,2,3,4,5,6,7},]
\addplot[color=red,mark=o,] coordinates {(1,50.39)(2,33.65)(3,12.38)(4,7.19)(5,6.09)(6,5.97)(7,5.89)};
\addplot[color=blue,mark=o,] coordinates {(1,51.14)(2,33.28)(3,15.75)(4,12.35)(5,12.10)(6,12.06)(7,12.08)};
\legend{(solar-30min),(solar-60min)}
\end{axis}
\end{tikzpicture}  
\caption{We compute the FNN Plot for the solar irradiance measurements at NTU Singapore for the year 2010.}
\label{fig:fnn-result}
\end{figure}
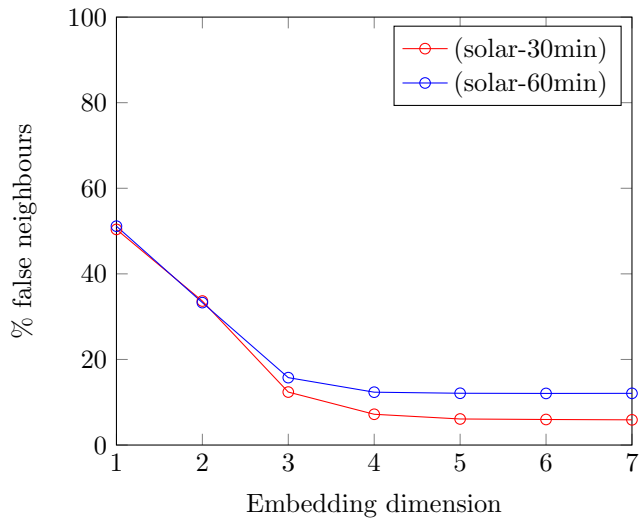

As we discussed in section \ref{embedding-dim}, the number of required previous data (input) for the forecasting is one less than the embedding dimension where the percentage of false neighbours saturates. Since we got 4 as the embedding dimension for both solar-60min and solar-30min, we need at least 3 previous data samples for the forecasting of both 60 minute resolution and 30 minute resolution time series data of solar irradiance.

The mean and standard deviation of solar-60min and solar-30min indicate the similarity between two data set. The result of false nearest neighbour method also shows similar patterns for these data set. Therefore, we can conclude that similar forecasting methods are suitable for 60 minute resolution data and 30 minute resolution data.

\psection{Conclusions and Future work}
In this paper, we have provided a chaotic analysis of solar irradiance recordings, measured at NTU Singapore. Our experiments on data sampled at $60$ minutes and $30$ minutes reveal an embedded dimension of four. This indicates that at least three previous measurements are necessary to reliably forecast a state of the solar irradiance measured at $60$ and $30$ minutes intervals. Our future work involves the use of higher temporal resolutions in such chaos theory analysis. We will also propose a framework for an accurate and reliable solar irradiance forecasting~\cite{dev2018solar,dev2016estimation}. 

\ack
The ADAPT Centre  for  Digital  Content  Technology  is  funded  under  the  SFI Research Centres Programme (Grant 13/RC/2106) and is co-funded under the European Regional Development Fund.

\bibliographystyle{IEEEbib}

\end{paper}

\end{document}